\documentclass[12pt,preprint]{aastex}

\begin{document}
\def\ss{SS~Cygni}
\def\ltorder{\mathrel{\raise.3ex\hbox{$<$}\mkern-14mu
             \lower0.6ex\hbox{$\sim$}}}

\shortauthors{MCGOWAN, PRIEDHORSKY \& TRUDOLYUBOV}
\shorttitle{X-RAY AND OPTICAL EVOLUTION OF SS~CYG}

\title{ON THE CORRELATED X-RAY AND OPTICAL EVOLUTION OF SS~CYGNI}
\author{K.E. McGowan\altaffilmark{1},
W.C. Priedhorsky\altaffilmark{1}, S.P. Trudolyubov\altaffilmark{1}}
\altaffiltext{1}{Los Alamos National Laboratory, MS D436, Los Alamos, NM 87545}
\email{mcgowan@lanl.gov}

\begin{abstract}

We have analyzed the variability and spectral evolution of the
prototype dwarf nova system \ss\ using {\em RXTE} data and
AAVSO observations.  A series of pointed {\em RXTE}/PCA observations
allow us to trace the evolution of the X-ray spectrum of \ss\ in
unprecedented detail, while 6 years of optical AAVSO and {\em
RXTE}/ASM light curves show long-term patterns.  Employing a technique
in which we stack the X-ray flux over multiple outbursts, phased
according to the optical light curve, we investigate the outburst
morphology.  We find that the 3 -- 12~keV X-ray flux is suppressed
during optical outbursts, a behavior seen previously, but only in a
handful of cycles.

The several outbursts of \ss\ observed with the more sensitive {\em
RXTE}/PCA also show a depression of the X-rays during optical outburst.
We quantify the time lags between the optical and X-ray outbursts, and
the timescales of the X-ray recovery from outburst.  The optical light
curve of \ss\ exhibits brief anomalous outbursts.  During these events
the hard X-rays and optical flux increase together.  The long-term
data suggest that the X-rays decline between outburst.
 
Our results are in general agreement with modified disk instability
models (DIM), which invoke a two-component accretion flow consisting
of a cool optically thick accretion disk truncated at an inner radius,
and a quasi-spherical hot corona-like flow extending to the surface of
the white dwarf.  We discuss our results in the framework of one such
model, involving the evaporation of the inner part of the optically
thick accretion disk, proposed by Meyer \& Meyer-Hofmeister (1994). 

\end{abstract}

\section{INTRODUCTION}
\renewcommand{\thefootnote}{\fnsymbol{footnote}}
\setcounter{footnote}{0}
\label{sect:intro}

\ss\ is a dwarf nova (DN) cataclysmic variable.  The main
characteristic of DN are outbursts which have recurrence timescales of
days to years, and last for days to weeks.  Outbursts occur in \ss\
every $\sim40$~d, when the source rises from a quiescent magnitude of 
$V\sim12$ to an outburst magnitude of $V\sim8.5$ \cite{szk84}.  The 
outbursts of \ss\ exhibit a bi-modal distribution, with wide and
narrow outbursts that last $\sim20$ and $\sim12$~d, respectively
\cite{szk84}. 

\ss\ was first detected in X-rays by Rappaport et al.\ (1974). The main
site of X-ray emission in DN is believed to be the boundary layer between 
the white dwarf and the accretion disk. Accretion material in the innermost 
orbit of the disk is decelerated in the thin boundary layer before coming 
to rest on the surface of the white dwarf. During this process the material 
should release $\sim$ half of its accretion luminosity.

\ss\ shows both hard and soft X-ray spectral components \cite{cord80}. 
The two spectral components are usually attributed to different states of 
the boundary layer flow, which depend on $\dot M$ \cite{prin79}. For 
$\dot M \ll \dot M_{\rm crit}$, $\dot M_{\rm crit}\sim 10^{16}$~g s$^{-1}$, 
the heated gas expands adiabatically and forms a hot ($T\sim 10^{8}$~K) 
corona which cools by thermal bremsstrahlung, radiating a hard spectrum. 
For $\dot M \gg \dot M_{\rm crit}$ the heated gas cools without
expanding out of the optically thick boundary layers, giving rise to
an approximate blackbody spectrum (at $T \sim 10^{5}$~K). During this 
high-accretion state, tenuous regions of the boundary layer flow,
above and below the center plane, remain in the hot, optically thin
state, yielding a residual hard component. 

Comparison of X-ray data from {\it EXOSAT} with the optical light curve 
of \ss\ by Jones \& Watson (1992) confirmed this X-ray behavior. In 
quiescence, the hard component is strong. The outburst is first seen in 
the optical band. After a short delay the hard X-ray component flares, 
then declines to a level below that of quiescence. The ultrasoft X-ray 
outburst begins as the hard X-ray flare ends. The ultrasoft X-ray emission 
tracks the optical emission closely during the peak of the optical burst, 
but the ultrasoft X-rays decline to quiescence more quickly. Once they start 
to decline, the hard X-ray component increases again, to a level brighter 
than quiescence, then declines more slowly than the optical in the
final decline.  Studies of detailed {\it RXTE}/PCA observations of
single outbursts of \ss\ exhibit this X-ray/optical behavior
(Wheatley, Mauche \& Mattei 2000; Priedhorsky, Trudolyubov \& McGowan
2002; McGowan, Priedhorsky \& Trudolyubov 2002; Wheatley, Mauche \&
Mattei 2003). 

We analyze several outbursts of \ss\ from archival data, to
investigate the optical and X-ray evolution of the source. In
particular, we want to confirm the optical/X-ray outburst
phenomenology found in previous works, which were based on studies of
only a few outbursts.

\section{OBSERVATIONS AND DATA ANALYSIS}
\label{sect:obs}

Its brightness and frequent outbursts makes \ss\ an ideal candidate
for observations by amateur astronomers.  Since its discovery in 1896,
\ss\ has been monitored by the American Association of Variable Star
Observers (AAVSO\footnote{http://www.aavso.org}).  We have obtained
the 1996--2001 light curve of \ss\ from the AAVSO. 

We have also obtained the X-ray light curve of \ss\ from the All Sky
Monitor (ASM; Doty 1988; Levine et al. 1996) that is onboard the 
{\it Rossi X-ray Timing Explorer (RXTE)}
satellite\footnote{http://xte.mit.edu} (Bradt, Rothschild \& Swank
1993).  The data contained in the archive includes background
subtraction.  The ASM has three X-ray channels spanning
$1.3-12.2$~keV\footnote{ftp://legacy.gsfc.nasa.gov}, from which we
derive $1.3-3.0$~keV ({\sc band1}) and $3.0-12.2$~keV ({\sc band2})
fluxes for our analysis.  While the ASM is small, with a $90$~cm$^{2}$
effective area, it is persistent in its coverage. 

We also use the much more sensitive, but sporadic, pointed observations 
taken by the Proportional Counter Array (PCA; Jahoda et al. 1996)
instrument on {\it RXTE} in the $3 - 20$ keV band. For our analysis we
used publicly available {\em RXTE}/PCA data collected during
$1996-2000$ (a total of $262$, observations $\sim 1000 - 20000$ s
duration). The observation log for the data used in our analysis is
shown in Table \ref{pca_obslog}.

The PCA data were reduced with the standard {\it RXTE} FTOOLS package, 
version 5.2.  We used PCA data collected in the $3 - 20$~keV energy range 
for the spectral analysis.  The PCA response matrices for individual 
observations were constructed using the {\em pcarmf} task and the background 
estimation was performed by applying {\em very large events} (VLE) or 
{\em faint} source models, depending on the level of the source flux 
\footnote{See http://heasarc.gsfc.nasa.gov/docs/xte/recipes/pcabackest.html}. 
The standard dead-time correction procedure was applied to the PCA data. In 
order to account for the uncertainties of the response matrix, a $0.5 \%$ 
systematic error was added in quadrature to the statistical error for each 
PCA energy channel. 

We generated energy spectra of \ss\ averaging the data over whole individual 
observations using the {\em Standard 2} mode data (binned data;
129-channel spectra accumulated every 16 s).  For the temporal
analysis of \ss\ we used the PCA data in {\em Standard 2} and {\em
Good Xenon} modes (unbinned data; time-stamped events with 256-channel
resolution).  

\section{COMPARISON OF AAVSO DATA AND ASM DATA}
\label{sect:opt_asm}

The contemporaneous AAVSO and ASM data cover the period from 1996
January 4 to 2001 August 30.  The data are shown in
Fig.~\ref{fig:opt_xr_lc}; while the optical outbursts are obvious, it
is difficult to see individual outbursts in the X-rays.

In order to investigate the optical and X-ray variability we have
adopted a stacking technique to average the optical and X-ray data and
improve signal to noise.  The analysis of the outburst and quiescent
(inter-outburst) data are presented in the following sections.

\subsection{Outburst}

\begin{figure}[h]
\vbox to5in{\rule{0pt}{5in}}
\includegraphics{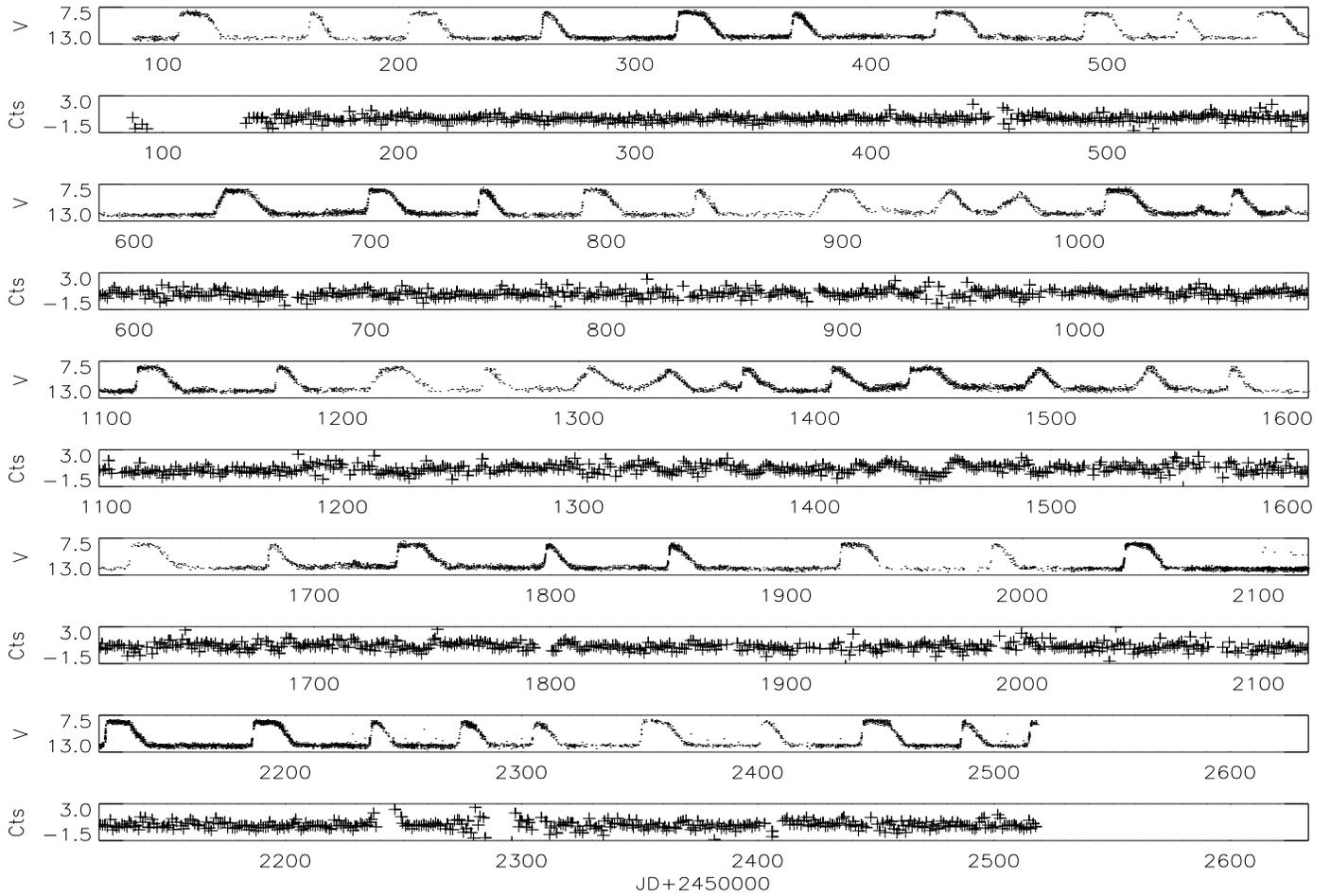}
\caption{Contemporaneous optical (panels 1, 3, 5, 7 and 9) and X-ray
(panels 2, 4, 6, 8 and 10) light curves of \ss\ from 1996 January 4 to
2001 August 30.  The optical data is from the AAVSO, the X-ray data is
from {\it RXTE}/ASM.  The X-ray data are binned in one day intervals.}\label{fig:opt_xr_lc}
\end{figure}

\begin{figure}[h]
\vbox to4in{\rule{0pt}{4in}}
\includegraphics{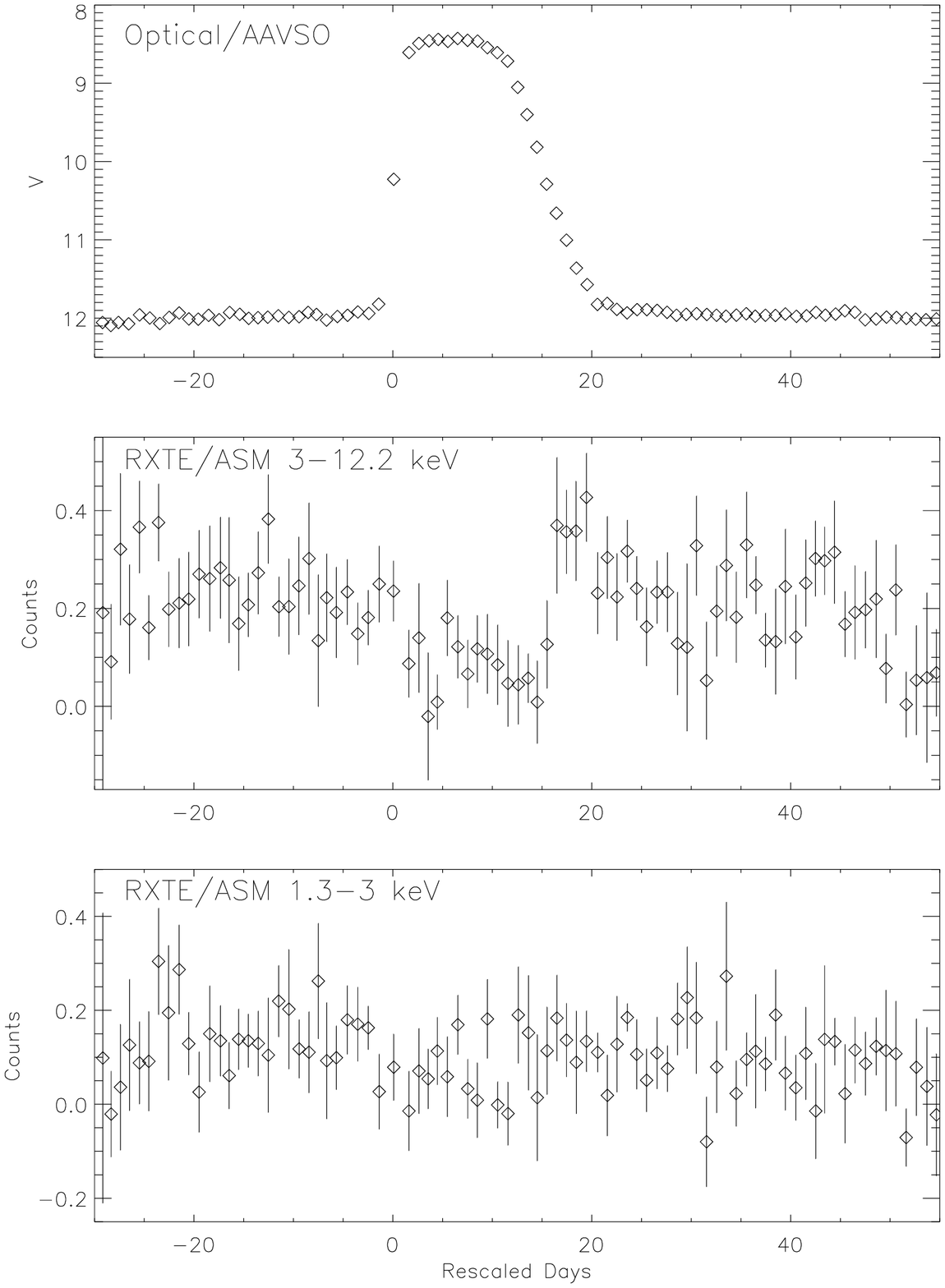}
\includegraphics{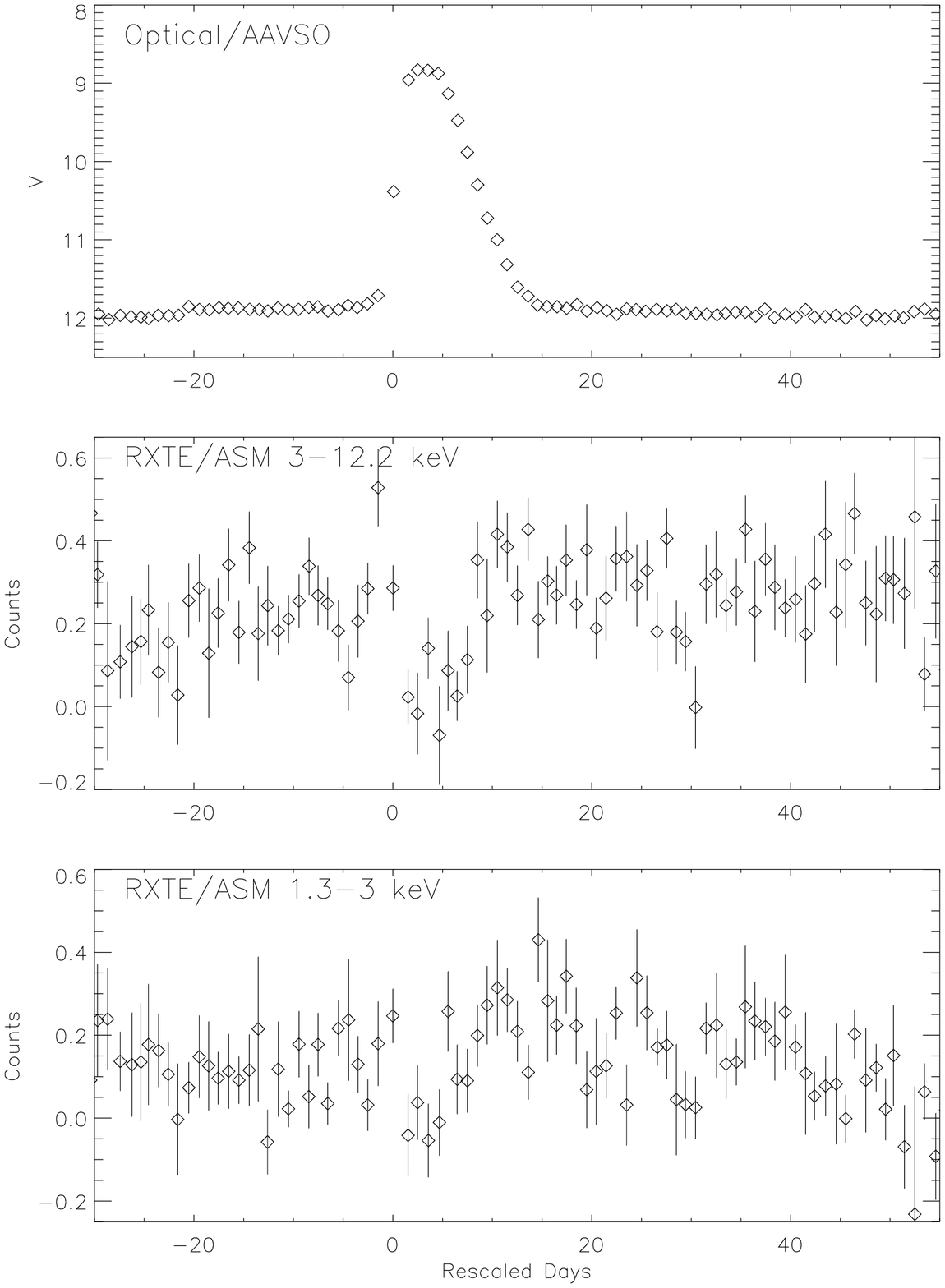}
\includegraphics{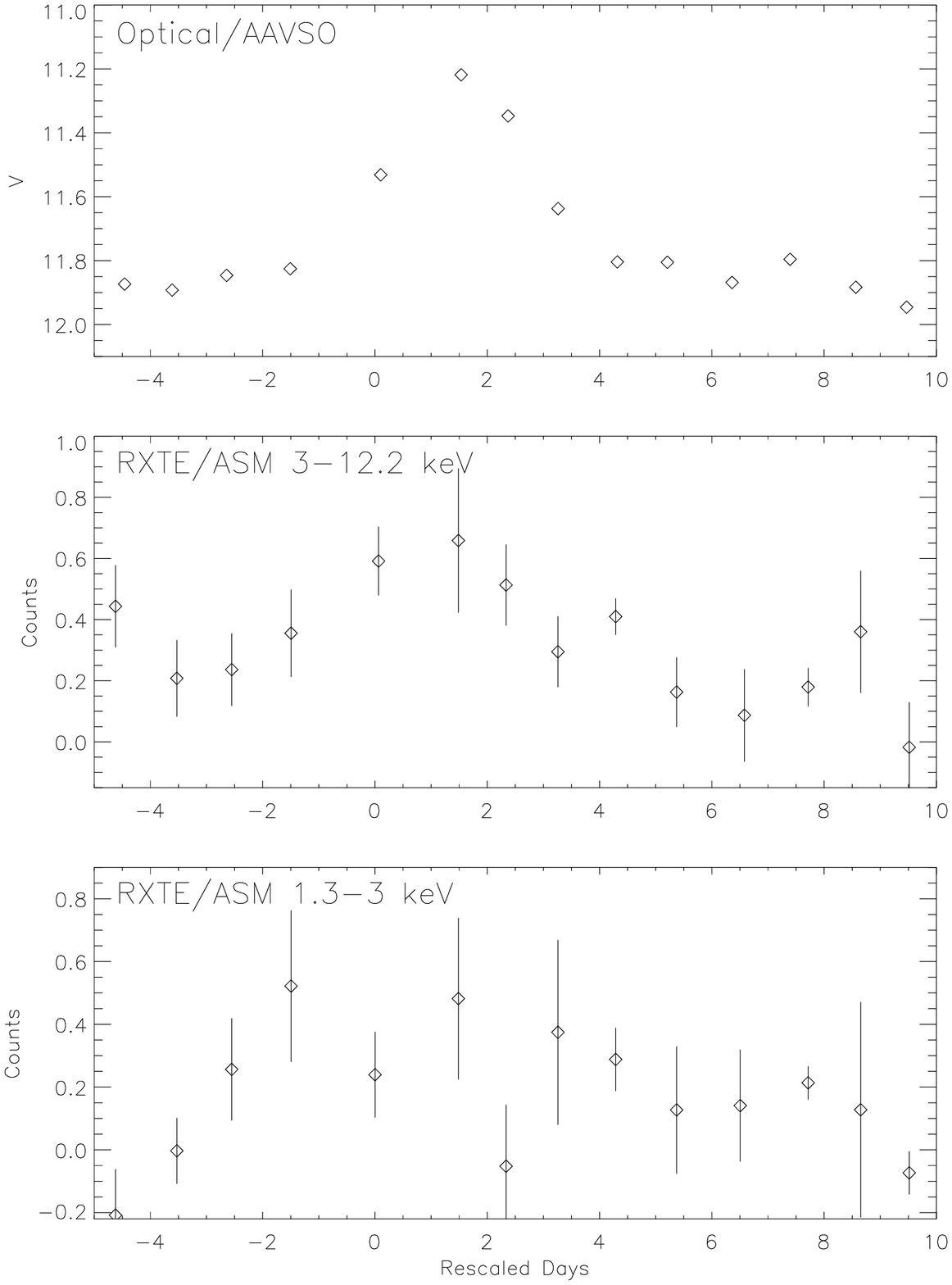}
\caption{Averaged outburst data of \ss\ produced using a stacking
technique (see text).  Data shown are for wide ({\it left}), narrow
({\it middle}) and anomalous ({\it right}) outbursts.  In each case,
the optical data is plotted in the {\it top} panel, the $3-12.2$ and 
$1.3-3$~keV X-ray data from the {\it RXTE}/ASM are plotted in the {\it
second} and {\it third} panels, respectively.}\label{fig:asm_outburst}
\end{figure}

We separate the optical data into wide (Fig.~\ref{fig:opt_xr_lc}, e.g.\
JD~2450320), narrow (Fig.~\ref{fig:opt_xr_lc}, e.g.\ JD~2450370), and 
anomalous or mini outbursts (Fig.~\ref{fig:opt_xr_lc}, e.g.\
JD~2451050).  In our data we find 22 wide, 27 narrow and 7 anomalous
outbursts.  For rescaling purposes we define start ($T_{start}$) and
end points ($T_{end}$) to occur at $V=10.5$ for the wide and narrow 
optical outbursts, and at $V=11.7$ for the anomalous outbursts.  By
calculating the average duration of the wide, narrow, and anomalous
outbursts we determine a characteristic length $\Delta T$ for the
three different outbursts ($\Delta T=16$~d for the wide, 9~d for the
narrow and 3~d for the anomalous outbursts).  To ensure we do not
ignore any data which may be part of an outburst we include $\sim30$~d
prior to $T_{start}$ and $\sim40$~d after $T_{start}$ in each cut of
data for the wide and narrow outbursts, and $\sim5$~d before and
$\sim10$~d after $T_{start}$ for the anomalous outbursts.  Using the
start and end times of the sections of optical data, we separate the 
{\sc band1} ($1.3-3.0$~keV) and {\sc band2} ($3.0-12.2$~keV) X-ray
data.  The optical and X-ray data are rescaled on the characteristic
outburst timescales such that $T_{0}=T_{start}$ and $T_{\Delta
T}=T_{end}$.  We can then calculate the average time history for each
type of outburst, in the optical, {\sc band1} and {\sc band2} X-rays,
by stacking and averaging the rescaled data
(Fig.~\ref{fig:asm_outburst}).    

For both the wide and narrow outbursts
(Fig.~\ref{fig:asm_outburst}, {\em left and middle panels}), the {\sc
band2} X-rays ($3-12.2$~keV) are anti-correlated with the optical
flux, i.e.\ {\em the {\sc band2} X-ray flux falls when the optical
flux rises during outburst}.  This is consistent with previous reports
that the hard X-rays are suppressed during optical outburst
\cite{JW92,pried02,mcgowan02,wheat03}.  We have confirmed that this
behavior holds over a large ensemble of outbursts.  The
previously-reported X-ray maximum at the decline of an outburst is
also visible.  The {\sc band2} X-rays for the narrow outburst shows an
increase in flux which could be associated with the
previously-reported X-ray maximum at the beginning of the optical
outburst (Fig.~\ref{fig:asm_outburst}, {\em middle panel}).  However,
as this is based on only one point, and a similar behavior is not seen
in the wide outburst, this result is inconclusive.  The averaged
anomalous outbursts (Fig.~\ref{fig:asm_outburst}, {\em right panel})
suggest a positive correlation between the optical and {\sc band2}
X-ray flux.    

\subsection{Quiescence}

\begin{figure}[h]
\epsscale{0.7}
\plotone{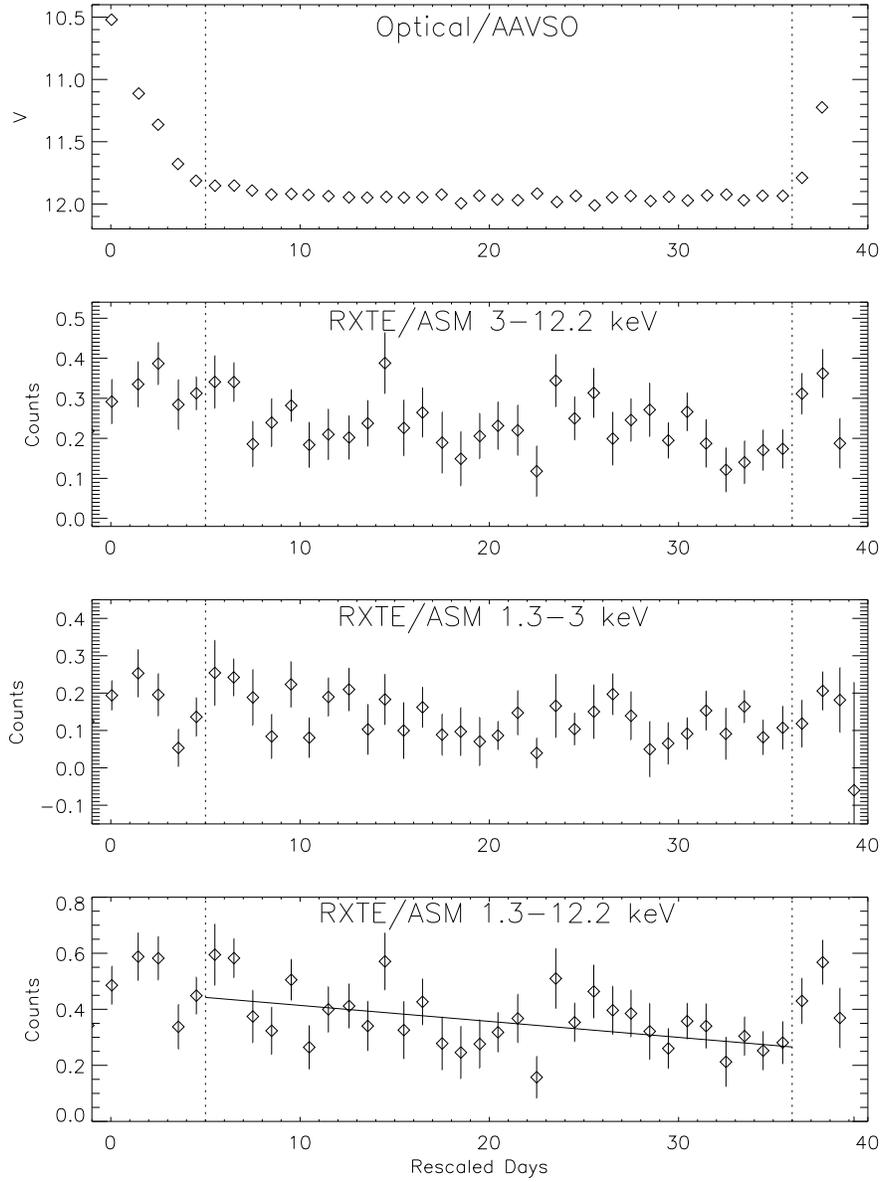}
\caption{Averaged data for the inter-outbursts.  {\it First} panel, optical
AAVSO data.  {\it Second} and {\it third} panels, $3-12.2$ and
$1.3-3$~keV ASM data, respectively.  Fourth panel, $1.3-12.2$~keV
ASM X-ray flux.  The {\it dotted} lines indicate the start and end points
of quiescence.  The {\it solid} line shows the best-fit decline to the 
data.}\label{fig:quies}
\end{figure}

We studied the optical and X-ray correlation during quiescence using a
similar method as for the outburst data.  We use the start and end
times of the optical outbursts determined above to separate the
inter-outburst data.  The inter-outburst was therefore the interval
between the $V=10.5$ crossing on the decline from outburst, and the
$V=10.5$ crossing on the rise to outburst.  We note that this is an
arbitrary definition of quiescence, which we use only for our stacking 
analysis.  We removed all data from times of anomalous outburst.
We calculate the average duration of quiescence, and define a
characteristic length for the quiescent interval of 38~d which we used
to rescale the data.  The optical, {\sc band1} and {\sc band2} X-ray
data are then averaged by stacking all the quiescent intervals
(Fig.~\ref{fig:quies}). 

A linear fit including a decline to the inter-outburst $1.3-12.2$~keV
X-ray flux of \ss\ results in a $\chi^{2}$ of 36.8 for 29 d.o.f.  We
chose the interval from rescaled days 5 to 36 as the largest span that
is clearly free from the optical outburst.  During this interval the
flux seems to fade, with a decline significant at the 3.6-sigma
level.  The error bars on the ASM X-ray fluxes reflect all currently
known systematic effects.  The best linear fit to the $1.3-12.2$~keV
X-ray flux between rescaled days 5 and 36 declines at the rate of
1.3\% per scaled day, with a decrease of $\sim40$\% over the day 5 --
day 36 interval (see Fig.~\ref{fig:quies}, {\it fourth panel}).  In
other words, the flux between days 5 and 36 was fit by a mean flux of
0.354 counts s$^{-1}$ and a slope of $-5.7 \times 10^{-3}$ counts
s$^{-1}$ day$^{-1}$, with an error in the slope of $\pm 1.6 \times
10^{-3}$ counts s$^{-1}$ day$^{-1}$, based on an increase in the
$\chi^{2}$ of 1 for one interesting parameter.  

\subsubsection{Folding on the Orbital Period}

\begin{figure}[h]
\vbox to4in{\rule{0pt}{4in}}
\includegraphics{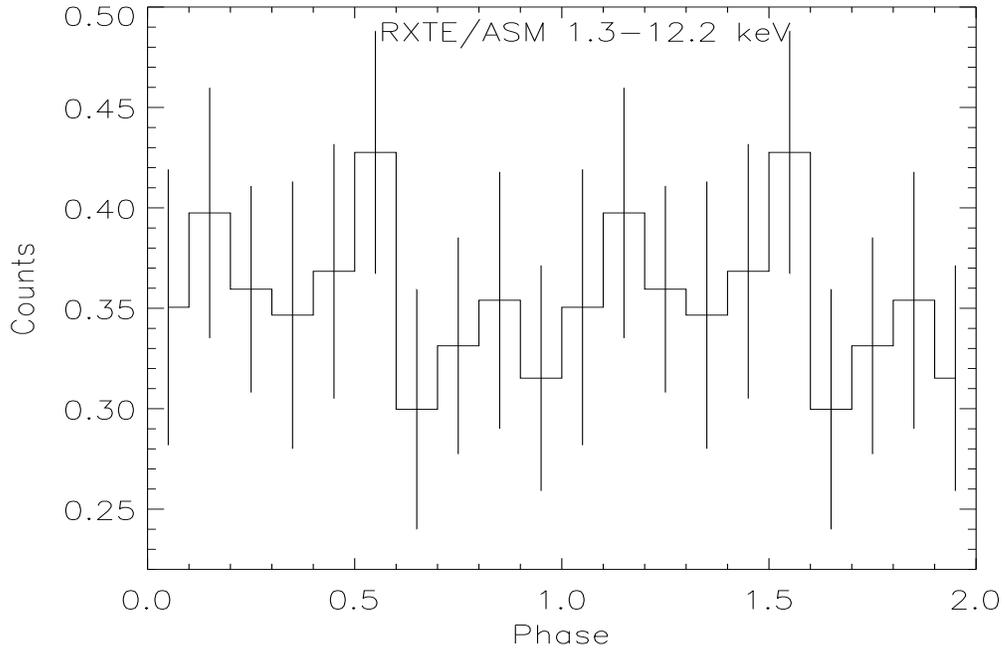}
\caption{Quiescent {\em RXTE}/ASM data in the $1.3-12.2$~keV range
folded on P$_{orb}$ = 0.2751297~d with T$_{0}$ = HJD 2447403.6295.}\label{fig:fold}
\end{figure}

To test for a modulation at the orbital period, we combined the {\sc
band1} and {\sc band2} X-ray data from the same quiescent intervals.
We folded the $1.3-12.2$~keV data on $P_{orb}=0.2751297$~d using 
$T_{0}=$ HJD 2447403.6295, where $T_{0}$ corresponds to the time of
inferior conjunction, i.e.~the time of zero velocity of the absorption
line star \cite{friend90}. The resulting light curve is
shown in Fig.~\ref{fig:fold}.   

We fit the data with a constant linear fit and a sinusoid.  The best
linear fit results in $\chi^{2}=7.1$ for 19 d.o.f., the best-fit
sinusoid gives $\chi^{2}=5.4$ for 16 d.o.f.  The semi-amplitude of the 
sinusoid is 0.024 counts which corresponds to a $7$\% modulation.
While the sinusoid gives a better fit to the data, employing the
$F$-test we find a probability of 78\%.  This indicates that the
variation is random, thereby failing to rule out a constant flux.

\section{COMPARISON OF SHORT-TERM OPTICAL AND X-RAY TIME HISTORIES}

\subsection{Light curve phenomenology. Correlated hard X-ray and 
optical evolution.}

\begin{figure}[h]
\vbox to4in{\rule{0pt}{4in}}
\includegraphics{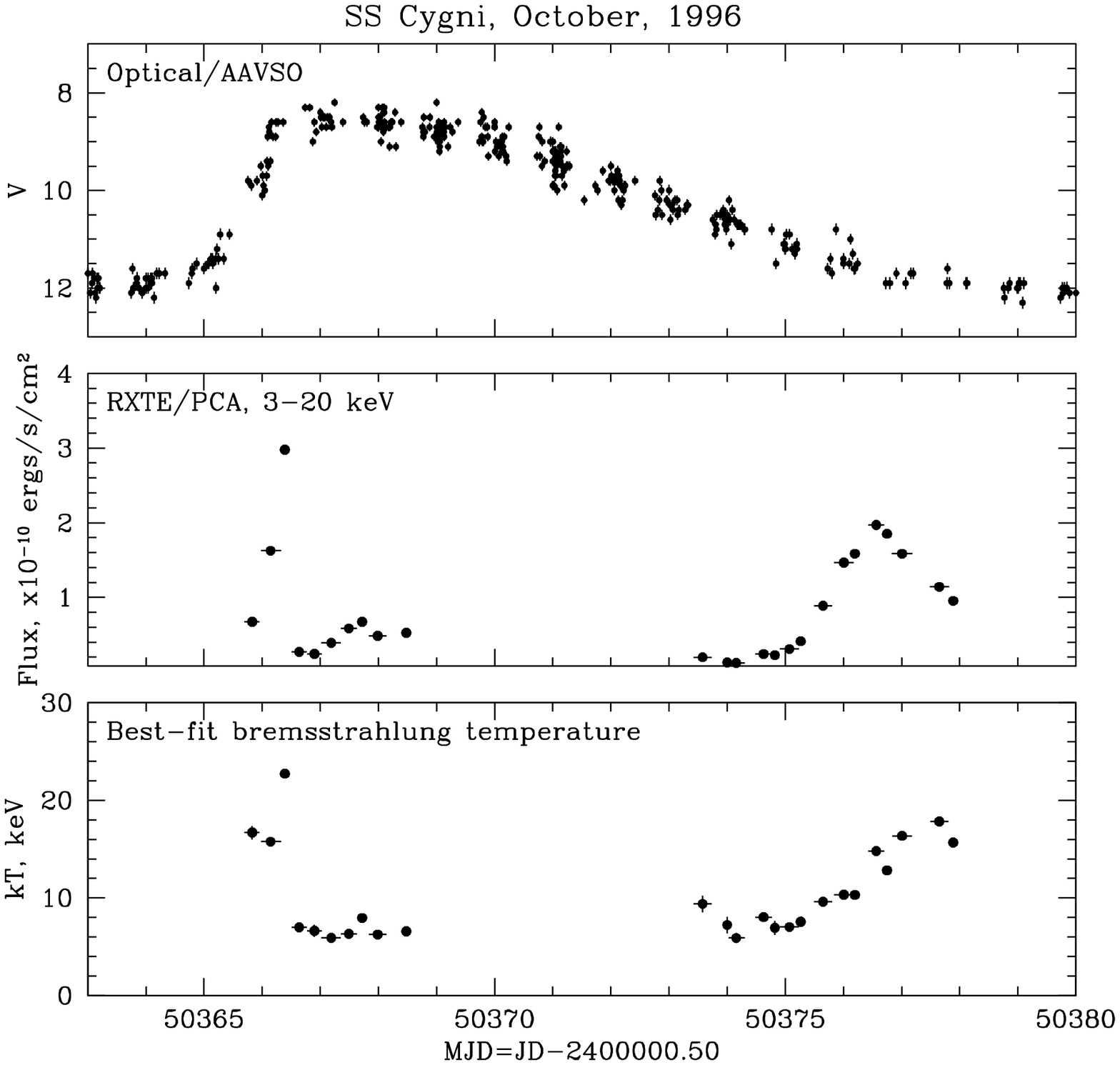}
\includegraphics{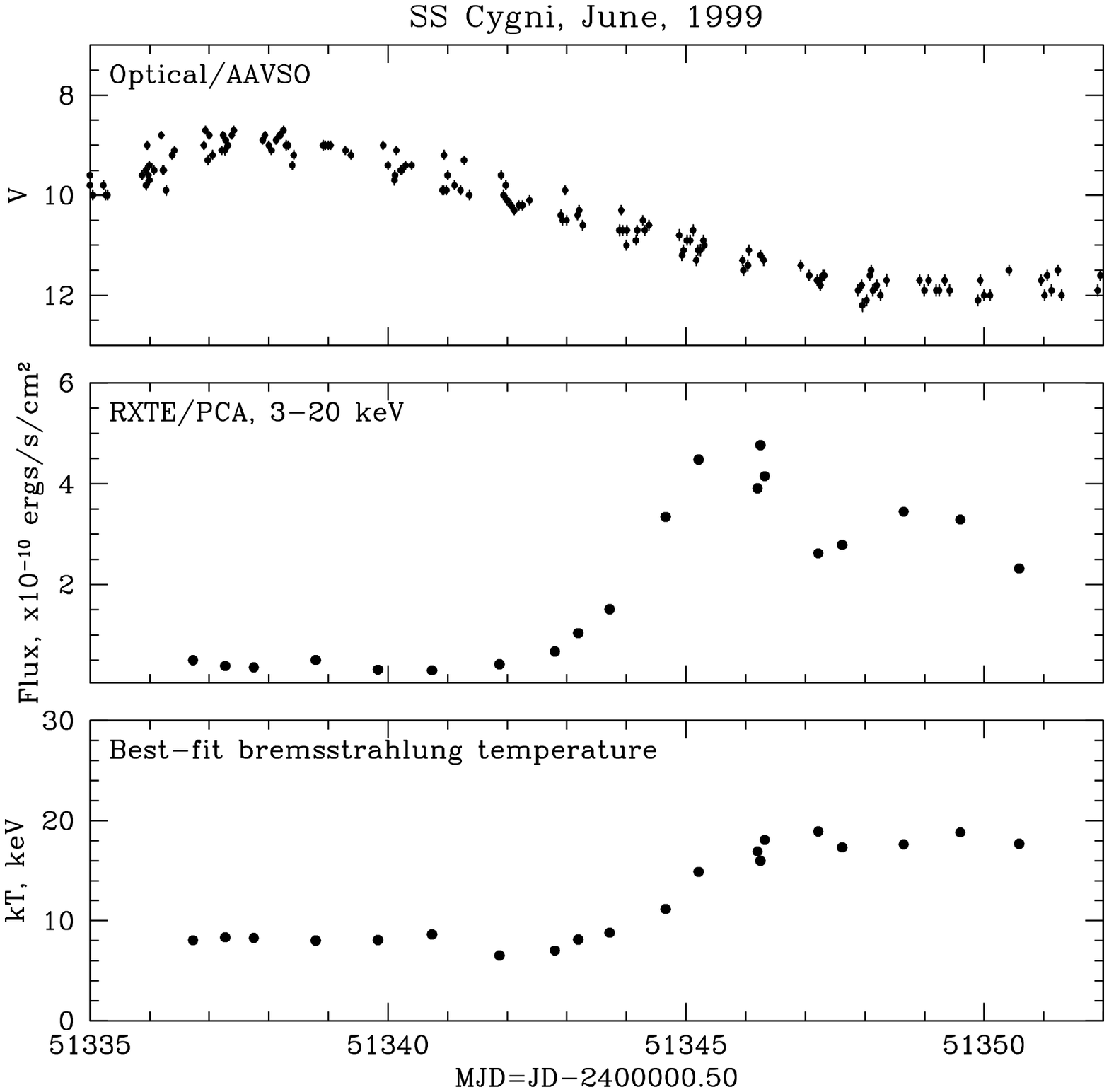}
\includegraphics{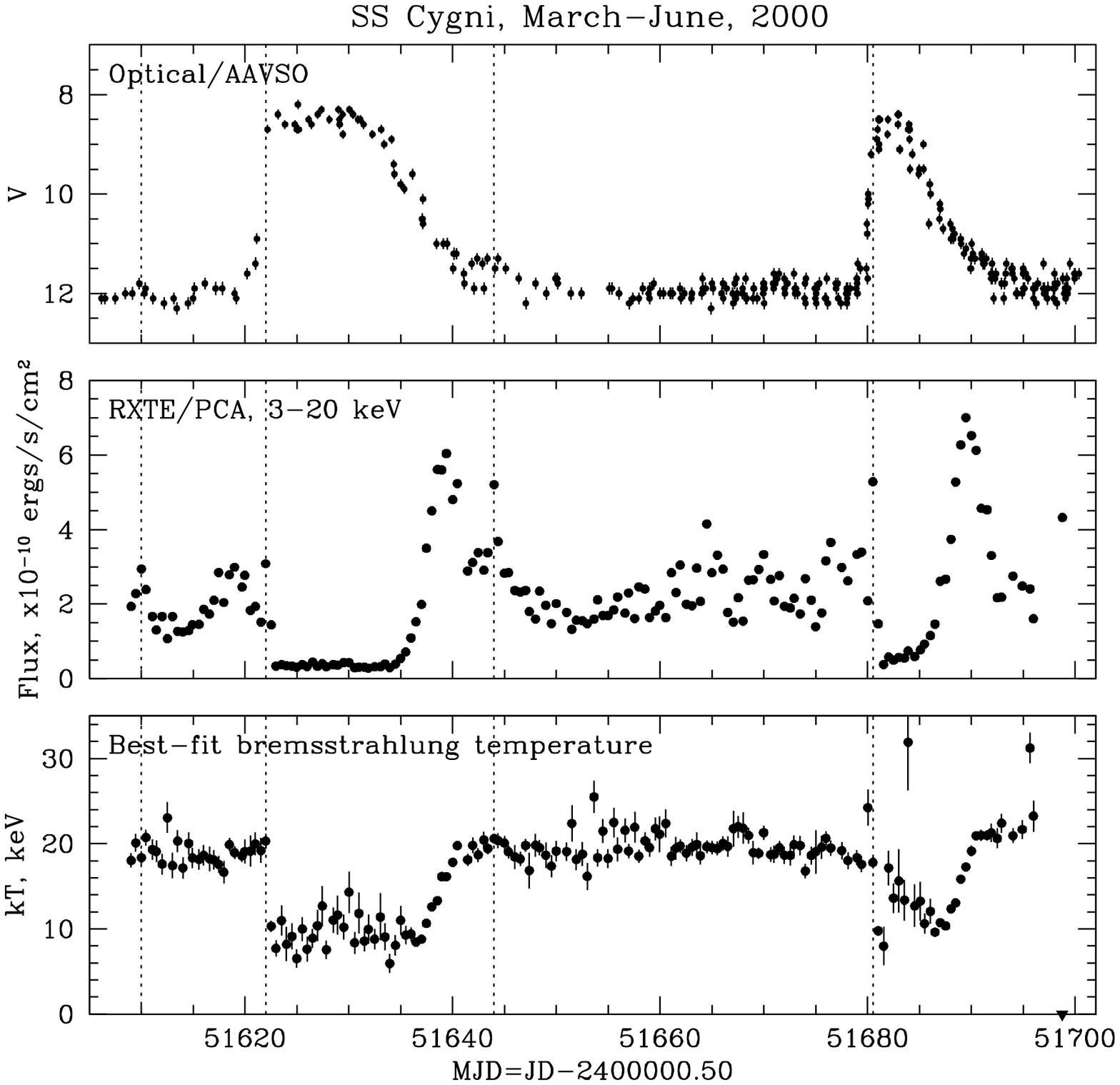}
\caption{Simultaneous optical (AAVSO) and X-ray ({\em RXTE}/PCA)
evolution of SS Cygni in October, 1996 ({\em left}), June, 1999 ({\em
middle}) and between March and June, 2000 ({\em right}).  The X-ray
data points represent the average of each of the individual PCA
observations which had exposure times ranging from
$\sim1000-20000$~s. The evolution of the optical and
X-ray flux are shown in the {\em upper} and {\em middle} panels. The
{\em lower} panel shows the evolution of the hardness of the X-ray
spectrum expressed in terms of the best-fit bremsstrahlung
temperature. Vertical {\em dotted} lines mark the positions of X-ray
{\it spikes}.}\label{fig:pca_outburst}
\end{figure}

In five years, six optical outbursts of \ss\ were covered with {\em
RXTE}/PCA pointed observations.  Two of the outbursts were poorly
sampled, therefore we present results from only four of the
outbursts here.  In spite of the differences in the duration of the
four outbursts, the morphology of the hard X-ray evolution ($3 -
20$~keV) is essentially the same in each (Fig.~\ref{fig:pca_outburst},
{\em upper and middle panels}).  Unlike the soft X-ray emission below
$\sim 1$ keV \cite{JW92}, the hard X-ray emission of SS Cygni shows a
general anticorrelation with optical flux, except for ``brief''
X-ray/optical outbursts ('spikes').  The positions of the {\em
spikes}, which last for $<2$~d, are marked in
Fig.~\ref{fig:pca_outburst} ({\em right panel}) at MJDs 51610, 51622,
51644 and 51680.  The rise of the optical outburst coincides with a
short {\em spike} in the hard X-rays: the flux increases by several
times, then fades, all in $\ltorder 1$~d.  At the maximum of the
optical outburst, the hard X-ray flux is at its lowest.  The decline
of the optical outburst corresponds to a gradual increase of the hard
X-ray flux, followed by a brief $2-4$~d maximum. Periods of optical
quiescence are generally characterized by a relatively high level of
hard X-ray flux. Rare anomalous optical outbursts are accompanied by
short X-ray {\em spikes} (i.e.\ MJDs 51610 and 51644 in
Fig.~\ref{fig:pca_outburst}, {\em right panel}).   

We studied the characteristic timescales of the hard X-ray luminosity 
transitions. The average duration of the X-ray {\em spike} coincident 
with an anomalous optical outburst is $\ltorder 2$~d, with a nearly 
symmetrical time history. Exponential fits to the rising/decaying parts 
of the X-ray {\em spikes} (MJDs 51610 and 51644) give an e-folding time 
of $\sim 0.5$~d (Fig.~\ref{fig:pca_outburst}). 

The X-ray {\em spikes} corresponding to the beginning of the optical 
outburst have a shorter duration of $\ltorder 1$~d. The observations of 
1996 October 9--11 allow a detailed study of the source evolution during 
such a {\em spike} (Fig.~\ref{fig:pca_outburst}, {\em left panel}). The light 
curve of \ss\ during the {\em spike}, based on four consecutive PCA 
observations \cite{wheat00,wheat03}, is shown in
Fig.~\ref{fig:ss_cyg_oct_1996_spike_lc}.  

The {\em spike} has an asymmetric shape with an initial quasi-exponential 
rise with $\tau_{rise} \sim 25000$~s followed by an abrupt decline in 
$\sim 5000$~s. The total duration of the hard X-ray {\em spike} is about half 
a day. The source $3 - 20$ keV flux rises by $\sim 5$ times from the quiescent 
level to the {\em spike} maximum and then drops by a factor of $\sim
18$ during the decline (Fig.~\ref{fig:pca_outburst}, {\em left
panel}). According to the AAVSO data, the optical outburst starts
$\sim 1$ day prior to the hard X-ray {\em spike} (Wheatley et al.\ 2000),
while the UV rise is coincident with sharp decline in hard X-rays.   

The hard X-ray recovery at the end of an outburst has a consistent pattern 
from one outburst to another (Fig.~\ref{fig:pca_outburst}). An exponential fit 
to the optical decline and X-ray rise is shown in Table
\ref{rise_phase_fits}.  The X-rays appear to evolve almost twice as
quickly as the optical light. 
  
\subsection{Energy spectra}

Previous X-ray studies of \ss\ \cite{swank79,cord80,JW92,yoshida92} 
have demonstrated the presence of two distinct spectral components: a hard 
component with a cut-off at $\sim 10 - 20$~keV (bremsstrahlung model), 
and an ultrasoft optically thick component that is cut-off at $\sim 60$ 
eV. These components behave very differently during optical outburst and
quiescence. The energy range of the PCA detector allows the direct study 
of the hard spectral component without contamination from an ultrasoft 
component. The ultrasoft component is invisible to the ASM, also.

\begin{figure}[h]
\epsscale{0.9}
\plotone{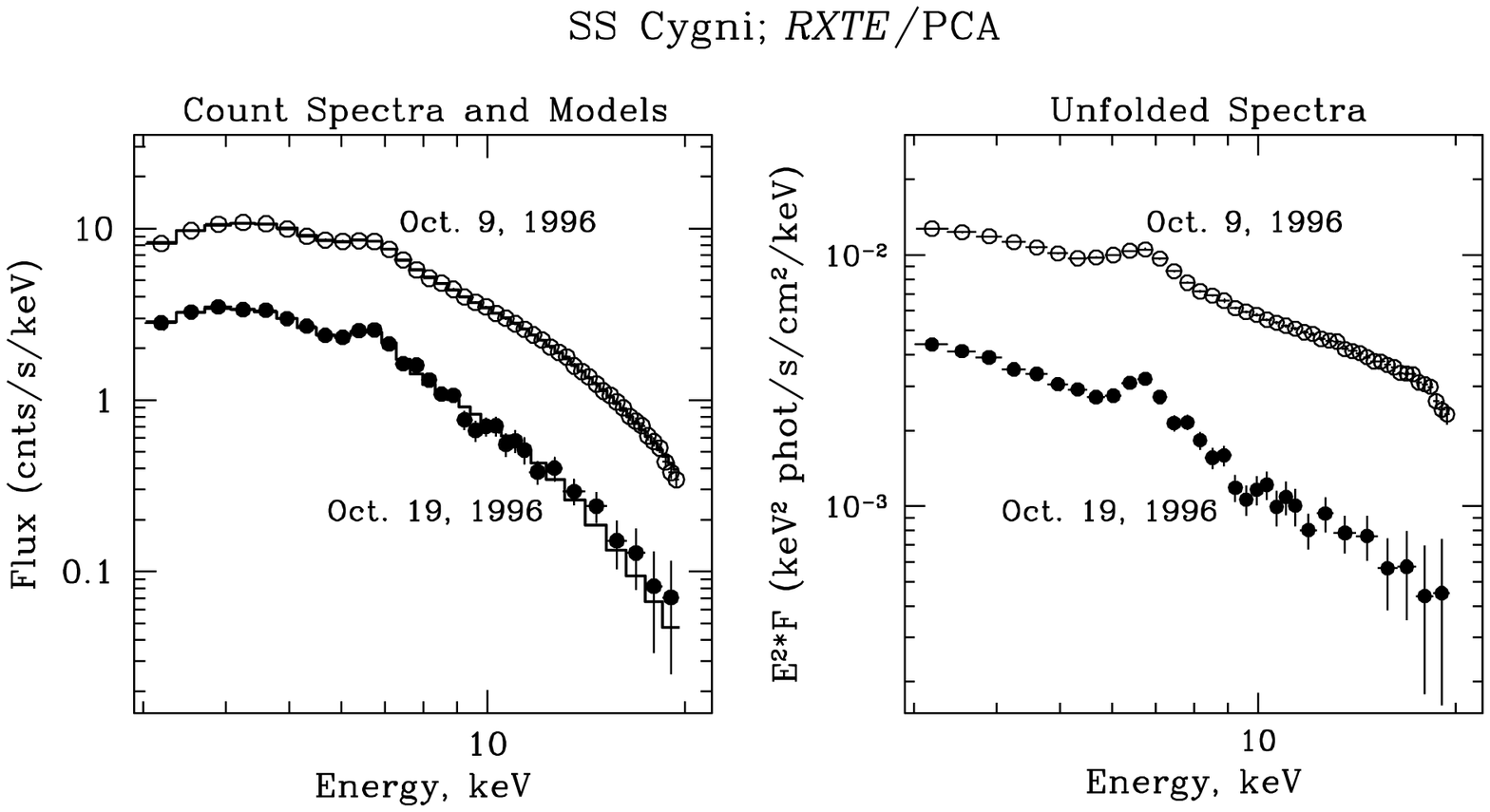}
\caption{Representative count ({\em left panel}) and corresponding 
unfolded ({\em right panel}) spectra of SS Cygni in the high (Oct. 
9, 1996) and low-flux (Oct. 19, 1996) states. PCA data, $3 - 20$ keV 
energy range. The best-fit analytical models (sum of bremsstrahlung 
spectrum and Gaussian emission line) are shown as histograms in the 
{\em left panel}. \label{fig:spec_general}}
\end{figure}

The {\em RXTE}/PCA spectral data were approximated with one of the
simplest models that fit the data: a two-component XSPEC (version 11)
model that includes an absorbed bremsstrahlung emission model and a
Gaussian emission line at $\sim 6.7$ keV.  Line emission from the
source can be present at 6.4, 6.7 and/or 6.9 keV, however the energy
resolution of the PCA is not sensitive enough to be able to
distinguish between these lines, and we are most likely seeing a
blend.  While this two-component model is unphysical, it allows us to
to compare our spectral results directly with previously published
results.  The width of the Gaussian line was fixed at 0.1 keV due to
the limited energy resolution of PCA detector ($\sim 1$ keV at 6 keV),
while the line centroid energy was left as a free parameter. Typical
energy spectra of \ss\ corresponding to the optical quiescence ({\em
upper histogram}) and outburst ({\em lower histogram}) are shown in 
Fig.~\ref{fig:spec_general}.

The combination of an absorbed bremsstrahlung radiation model and a 
Gaussian emission line at $\sim 6.7$~keV provides a satisfactory
description for the PCA spectral data in the $3 - 20$ keV energy band
(Fig.~\ref{fig:spec_general}).  The best-fit value of the
characteristic temperature, $\rm kT_{\rm bremss}$, of the
bremsstrahlung model varies between $\sim 6$ and $\sim 26$~keV, and
the resulting model $3 - 20$ keV energy flux changes from $\sim
10^{11}$ to $\sim 7 \times 10^{-10}$ ergs s$^{-1}$ cm$^{-2}$. The
evolution of best-fit spectral parameters (X-ray flux and $\rm kT_{\rm
bremss}$) during the 1996, 1999 and 2000 {\em RXTE}/PCA observations
is shown in Fig.~\ref{fig:pca_outburst} ({\em middle} and {\em lower}
panels). There is a unique correlation between both the hard X-ray
flux and hardness of the X-ray spectrum, and the optical flux level.
The X-ray flux is high and the X-ray spectrum is hard during optical
quiescence, and at short periods at the beginning and the end of
optical outburst. On the other hand, during the maximum of optical
outburst the X-ray flux is extremely low and the X-ray spectrum is
soft (corresponding to the lowest level of cut-off energy). The
relative strength of the 6.7 keV emission line also is correlated with
optical flux: the equivalent width of the emission line rises from
$\sim 350$~eV during optical quiescence to $\sim 800$~eV during the
maximum of the optical outburst.  

The relation between the level of X-ray flux and the best-fit value of 
$\rm kT_{\rm bremss}$ (hardness-intensity diagram) is shown in 
Fig.~\ref{fig:hardness_flux}. Each point represents the data averaged over one 
individual observation. Fig.~\ref{fig:hardness_flux} clearly shows that the 
observations can be subdivided into three distinct regions according to 
their positions on the hardness-intensity diagram, illustrating the 
correlation between X-ray and optical evolution.

The first, densely populated region is concentrated around a temperature 
of $\sim 20$~keV and an X-ray flux of $\sim 2 \times 10^{-10}$ ergs s$^{-1}$ 
cm$^{-2}$. This region corresponds to the periods of optical quiescence, 
with a relatively high level of hard X-ray flux (Fig.~\ref{fig:pca_outburst}). 
The second, densely populated region is concentrated around a temperature 
of $\sim 8$~keV and an X-ray flux of $\sim 3 \times 10^{-11}$ ergs s$^{-1}$
cm$^{-2}$. This group represents the periods of optical/soft X-ray outburst, 
characterized by an extremely low level of hard X-ray flux 
(Fig.~\ref{fig:pca_outburst}). The third, sparsely populated group comes
mostly from periods of transition between the optical outburst and
quiescence, characterized by the rise of hard X-ray luminosity and
general hardening of X-ray spectrum (Fig.~\ref{fig:pca_outburst}). A small
fraction of the observations from the third group correspond to the
fast transitions from optical quiescence to the outburst.

\section{DISCUSSION}
\label{sect:dis}

It is generally believed that outbursts in DN are caused by a sudden
brightening of the optically thick accretion disk due to an increased
accretion rate (see Osaki 1996 for review). Two models were proposed to
account for the intermittent accretion, the mass-transfer instability
model (MTI; Bath 1973) and the disk instability model (DIM; Osaki
1974). It is now generally accepted that the DIM is a strong
factor for DN outbursts \cite{osaki96}. In the DIM the secondary
star's mass transfer rate is continuous, but during quiescence the
mass flow from the secondary is not fully accreted onto the white
dwarf but is mostly held up in the outer parts of the accretion
disk. Once the stored mass reaches a critical amount, a thermal
instability \cite{meyer81} within the disk causes this matter to
flow through the disk and onto the white dwarf, resulting in optical
and UV/soft X-ray outburst.  

The general pattern of dwarf novae optical and X-ray behavior during
outburst is traditionally explained in the framework of the DIM
\cite{lasota01}. Two observed features, however, appear to pose
significant problems for the simple single component disk models
\cite{meyers94} which are unable to reproduce: i) a sizable lag in
rise-time of the UV light with respect to optical and hard X-rays rise
at the beginning of the outburst and ii) the appearance and gradual
decline of hard X-rays and UV light when the system falls back into
quiescence. 

The observed optical/hard X-ray/UV delay and the post-outburst
evolution of the dwarf novae can be explained by considering models in
which the inner edge of the accretion disk does not extend down to the
surface of the white dwarf.  Truncation of the disk can be due to
evaporation of the inner part of the accretion disk \cite{meyers94},
magnetic fields (e.g.~Livio \& Pringle 1992), or to an
advection-dominated accretion flow (ADAF; Menou 2001).  Here we
examine the observed multiwavelength behavior of \ss\ in the framework
of the first model, evaporation of the inner disk.  Let us assume that
the quiescent accretion flow in \ss\ consists of an outer optically
thick cool disk truncated by evaporation at a certain radius and,
inside that, an optically thin quasi-spherical boundary layer/corona
structure extending down to the surface of the white dwarf.  

\begin{figure}[h]
\epsscale{0.8}
\plotone{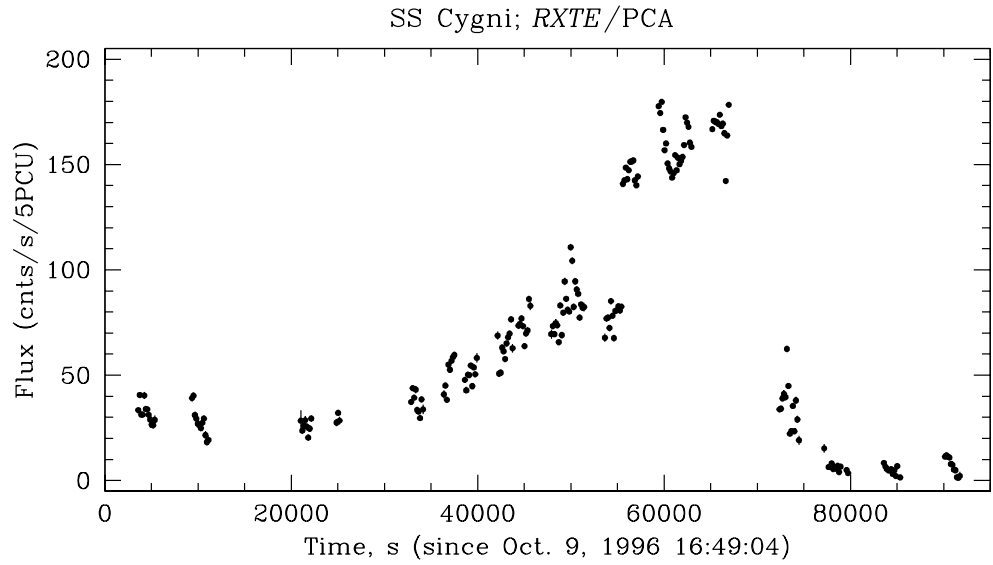}
\caption{A detailed evolution of the X-ray flux from SS Cygni during the
X-ray {\em spike} on Oct. 9 -- 10, 1996 (MJD $\sim 50366$, see also
{\em left panel} in Fig. \ref{fig:pca_outburst}). {\em RXTE}/PCA
data, $3 - 20$ keV energy range. \label{fig:ss_cyg_oct_1996_spike_lc}} 
\end{figure}

The relative timing of the optical, hard X-ray, and soft X-ray/UV
components in an \ss\ outburst are now well established
\cite{wheat03}.  We suggest that this sequence is an important clue
for understanding the movement of material through the disk.  To
summarize the outburst phenomenology, after the initial optical
decline, there is a short delay ($\sim 0.5$ days) before the flaring
of the hard X-ray component and an additional $\sim 1$ day delay
before the flare in the soft X-ray and UV bands
\cite{swank79,wheat00,wheat03}.  The key features of the hard X-ray
flare are the fast quasi-exponential rise of the flux ($\tau_{rise}
\sim 25000$~s) and an abrupt decline
(Fig.~\ref{fig:ss_cyg_oct_1996_spike_lc}) coincident with the rise of
the UV flux (Wheatley et al.\ 2000). 

These delays are conventionally interpreted as that passage of a
heating wave through the accretion disk (Wheatley et al.\ 2003 and
references therein). The wave starts in the outer part of the disk,
causing the optical rise, then reaches the innermost region, where it 
initiates an increase in the mass flow from the inner edge of the disk
to the boundary layer and white dwarf. This gradual increase in flow
first yields an increase in the optically thin, hard X-ray emission;
at a certain critical point, the flow switches its mode into an
optically thick, soft X-ray emitting plasma.

We propose an alternative explanation for the optical/hard X-ray/
soft X-ray sequence, inspired by the fact that the optical/hard X-ray
delay is comparable to the dynamical timescale to fall in from the
outer region of the accretion disk. In our picture, when the outburst
starts, the outer disk transitions from a cool to a hot state.  The
increased mass accretion rate gives rise to the optical flare.  The
increase of this mass transfer rate, either in the outer part of the
accretion disk, or in the vicinity of the inner Lagrangian point,
somehow leads to an increase of the mass flow through the optically
thin inner region of the accretion flow on a dynamical timescale that
is comparable to the orbital period of the \ss\ system.  This requires
a path for the accreting material that bypasses diffusive flow through
the disk, perhaps via the sub-Keplerian flow above the disk.  Once
this material arrives at the inner region, it causes the initial fast
rise of the hard X-ray flux (Fig.~\ref{fig:ss_cyg_oct_1996_spike_lc}).
The accretion rate through the optically thick accretion disk
increases on a diffusive timescale, causing the rise of UV and soft
X-ray flux.  At some point, the rise of the UV/soft X-ray flux,
combined with the increase of the accretion rate through the optically
thin region, will lead to its catastrophic cooling and an abrupt
softening of X-ray spectrum and abrupt drop of hard X-ray flux, as
seen in the {\em RXTE} light curve
(Fig.~\ref{fig:ss_cyg_oct_1996_spike_lc}).  

In this alternative picture, the optically thick disk/boundary layer
will extend down to the white dwarf surface when the outburst is at
maximum, and radiation driven disk winds will probably strip the
coronal layers from the disk \cite{meyers94}.  The decline of the
accretion rate during late stages of the outburst allows disk
evaporation to resume, restoring the optically thin disk corona.  As
the corona returns, the hard X-ray flux recovers and the X-ray
spectrum hardens.  At the end of the optical outburst we expect that
the hard X-ray flux should decay more slowly than the optical, as we
observe, because the variation in $\dot M$ will be ``smeared out'' as
it travels through the disk \cite{bath81,JW92}.  The hot/cool
transition seen directly in the optical is thereby delayed and spread
out as it reaches the innermost disk.  

This picture also explains the phenomenon of short hard X-ray spikes
associated with optical anomalous outbursts.  The short-term increase
of the mass transfer rate in the outer optically thick disk or in the
vicinity of the inner Lagrangian point creates an increase of the mass
flow in the inner optically thin region, by the same bypass that takes
place in a normal outburst, and causes an increase in the hard X-ray
flux.  However, the smaller short spikes decays as the small mass
spike is accreted, rather than being quenched by the optically
thick flow through the diffusive disk as in a normal outburst. 

\begin{figure}[h]
\epsscale{0.8}
\plotone{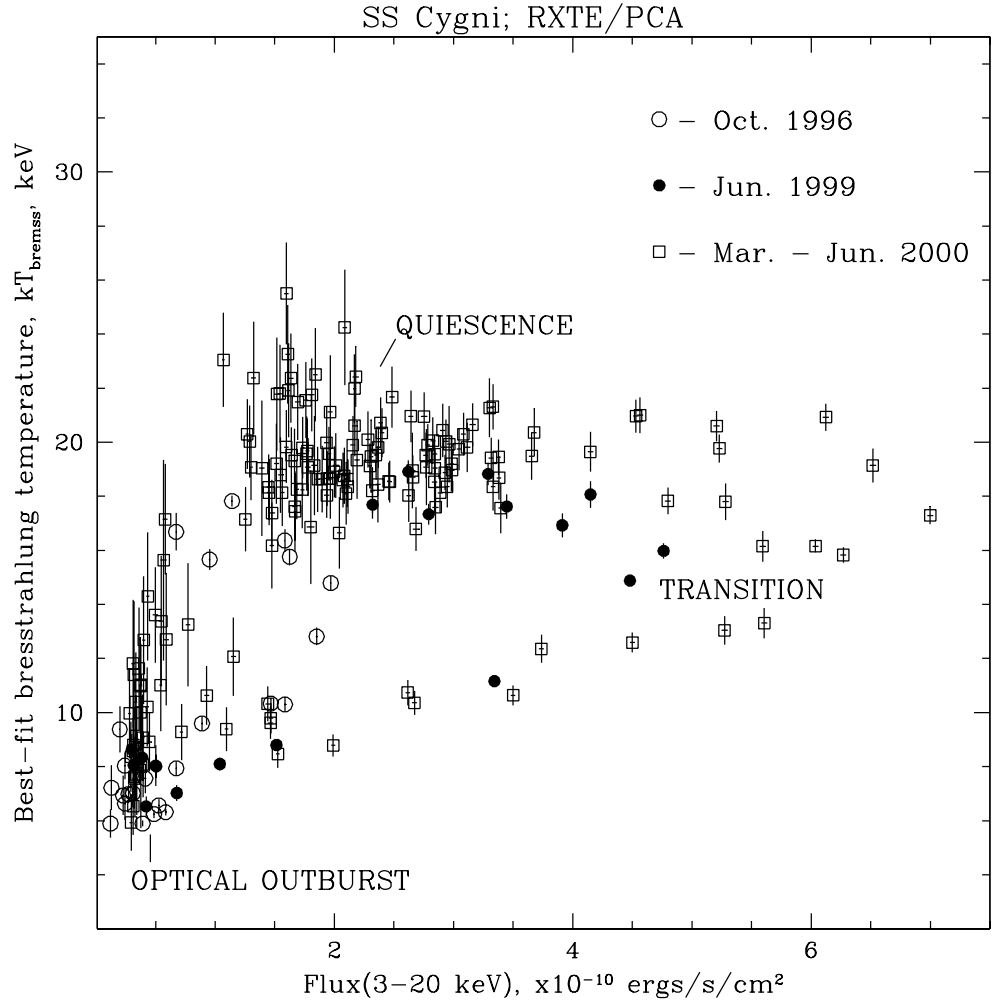}
\caption{Hardness of the X-ray spectrum in terms of best-fit bremsstrahlung 
temperature vs. total X-ray flux in the $3 - 20$ keV energy range.
The {\em open} and {\em filled} circles represent 1996, October and
1999, June data, {\em open} squares correspond to March-June, 2000
observations. \label{fig:hardness_flux}}
\end{figure}

The series of pointed {\em RXTE}/PCA observations allow us to trace
the evolution of the X-ray spectrum of \ss\ in unprecedented detail.
The results of spectral analysis of {\em RXTE}/PCA observations of
\ss\ provide additional support for the two-component model of the
accretion flow mentioned above. The hard X-ray emission from
non-magnetic CVs can be interpreted as due to: i) optically thin
radiation from strong shocks formed in the boundary layer when the
mass transfer rate was low; ii) an X-ray emitting corona-like
structure above the plane of the disk/boundary layer
(Meyer \& Meyer-Hofmeister 1994; Frank, King \& Raine 2002).  In
quiescence and at low disk accretion rates, the optically thin
boundary layer gas is likely to be thermally unstable to turbulent
viscous heating, rapidly achieving hard X-ray temperatures ($\sim
10^{8}$ K).  At the same time, this gas is likely to expand out of the
disk and boundary layer, and form a hot corona-like structure around
most of the white dwarf.  At quiescence (low $\dot{M}$) \ss\ shows
relatively bright, highly variable hard X-ray emission (see Wheatley
et al.\ 2003).  The energy spectrum of the source is extremely hard
with a cut-off at $\sim 20 - 25$ keV (corresponding to the
``QUIESCENCE'' and upper parts of the ``TRANSITION'' region in
Fig.~\ref{fig:hardness_flux}) indicating the temperature of the
optically thin emitting gas is $\gg 10^{7}$ K.  At high disk accretion
rates (the maximum of the optical outburst) the instability will be
largely suppressed, and the boundary layer luminosity mostly emitted
as soft X-rays, while the corona-like structure will be completely
cooled down by rising flux of soft photons from the disk/boundary
layer. After this stage we can expect the radiation driven winds to
remove the coronal layer, and the hard X-rays observed during optical
outburst will originate from shocks connected with such winds
\cite{lucy70,mauche87}. The X-ray spectrum of \ss\ during the optical
outburst is much softer than in quiescence, with a cut-off energy of
$\sim 6 - 8$ keV (Fig.~\ref{fig:pca_outburst},
\ref{fig:hardness_flux}). The abrupt softening of the X-ray spectrum
observed with {\em RXTE}/PCA at the end of the hard X-ray spikes
(Fig.~\ref{fig:pca_outburst}, \ref{fig:ss_cyg_oct_1996_spike_lc}) also
supports the above interpretation. The recovery of the hard X-ray flux
during the decline of the optical outburst is accompanied by gradual
hardening of the source energy spectrum
(Fig.~\ref{fig:pca_outburst}, \ref{fig:hardness_flux} (the
``TRANSITION'' region)), as expected from the recovery of the inner
hot optically thin region. The subsequent decline of the optical and
hard X-ray luminosity can be attributed to the further decrease of the
mass accretion flow through the optically thick and thin regions of
the accretion disk. 
      
Our analysis of the {\em RXTE}/ASM and AAVSO observations of \ss\ over
a period of 6 years confirms the general picture of hard X-ray
evolution deduced from more sensitive pointed observations of a
handful of individual outbursts
\cite{swank79,cord80,JW92,yoshida92,wheat00,wheat03}. The
average ASM light curves of \ss\ obtained by stacking a large number of
individual outbursts still show the suppression of the hard ($3 -
12.2$ keV) X-rays during both wide and narrow optical outbursts
(Fig.~\ref{fig:asm_outburst}).        

The {\em RXTE}/ASM observations of \ss\ during quiescence show some
evidence for a decrease in the hard X-ray flux during the
inter-outburst periods.  The observations of the post-outburst
evolution of other dwarf novae often show a decrease of the optical
and UV flux between outbursts \cite{smak00}.  The observed decrease in
the X-ray, UV and optical poses additional problems for the simple DIM
models, which imply a gradual increase of the luminosity (in all
bands) during quiescence. On the other hand, the two-component model
of the accretion flow involving disk evaporation \cite{meyers94}
naturally solves this problem.  As the outburst approaches its end,
the inner edge of the optically thick accretion disk progressively
moves outwards with time.  The accretion through the coronal flow
thereby decreases, causing a decline in the hard X-ray flux
\cite{meyers94}.  This decrease in X-rays reduces the level of X-ray
flux illuminating the outer parts of the accretion disk.  Hence, less
X-rays are reprocessed which leads to a decrease in the optical flux.
   
\section{CONCLUSIONS}
\label{sect:conc}

We have analyzed the variability and spectral evolution of the prototype
dwarf nova system \ss\ using simultaneous {\em RXTE} data (hard
X-rays) and AAVSO (optical) observations. The series of pointed {\em
RXTE}/PCA observations allow us to trace the evolution of the X-ray
properties of \ss\ in unprecedented detail. We find that our
results are in general agreement with predictions of modified disk
instability models that imply a two-component accretion flow, formed
through the evaporation of the inner part of the optically thick
accretion disk \cite{meyers94} or other process. We find that the
optical and X-ray outburst morphology of \ss\ predicted by the
modified DIM and observed in a series of pointed observations is
confirmed by stacking the marginal detections of individual outbursts
from {\em RXTE}/ASM observations. The averaged light curves for the
wide and narrow outbursts display a suppression of the hard ($3 -
12.2$ keV) X-ray flux during optical outbursts.

Based on the hard X-ray ({\em RXTE}/PCA/ASM) data and optical
(AAVSO) observations, we quantify the time lags between the optical
and X-ray outbursts, and the time scales of the X-ray spectral
recovery from outburst. We also find that anomalous optical outbursts
of \ss\ are accompanied by short hard X-ray outbursts lasting $\sim 1 - 2$ 
days. 

Using the {\em RXTE}/ASM data, we find evidence for a decrease in the
hard X-ray flux during quiescence. While this is in disagreement with
a simple DIM which predicts an increase in the disk mass, constant or 
decreasing flux has been observed in optical and UV studies of other
dwarf novae (e.g.\ Smak 2000). It should be mentioned that a modified
DIM involving a two-component accretion flow \cite{meyers94}
naturally solves this problem, by predicting the decrease of hard
X-ray and optical/UV flux due to the outward motion of the inner
boundary of optically thick accretion disk due to its evaporation into
a corona-like structure.  
   
We folded the quiescent $1.3-12.2$~keV X-ray data on the orbital period,
but do not find any significant modulation.  An all-sky monitor a few
times more sensitive than the {\em RXTE}/ASM would allow us to study
individual outbursts and the quiescent behavior of \ss\ with much
greater signal to noise, and to extend these studies to a large number
of dwarf nova (Priedhorsky, Peele \& Nugent 1996).  

\section{ACKNOWLEDGMENTS}

We thank Peter Wheatley for useful discussions. This research has made
use of data obtained through the High Energy Astrophysics Science
Archive Research Center Online Service, provided by the NASA/Goddard
Space Flight Center. In this research, we have used, and acknowledge
with thanks, data from the AAVSO International Database, based on
observations submitted to the AAVSO by variable star observers
worldwide.

\clearpage
\begin{table*}
\begin{center}
\small
\caption{Pointed {\em RXTE}/PCA observations of SS Cygni 
used in this analysis. \label{pca_obslog}}
\begin{tabular}{ccccc}
\hline
\hline
Proposal ID & Year & Start Date -- End Date & No. of observations &
Total exp. time (ks) \\
\hline
P10040 & 1996 & Oct. 9 -- Oct. 21& $25$  & 500 \\
P20033 & 1997 & Mar. 1 -- Jul. 2 & $42$  & 190 \\
P40012 & 1999 & Jun. 7 -- Jun. 21& $20$  &  70 \\
P50011 & 2000 & Mar. 5 -- Jun. 3 & $185$ & 147 \\
\hline
\end{tabular}
\end{center}
\end{table*}

\begin{table*}
\begin{center}
\small
\caption{Parameters of an exponential approximation for the rising phase 
in the $3 - 20$ keV energy band ({\em RXTE}/PCA data) and decline in
the optical band (AAVSO data). \label{rise_phase_fits}}
\begin{tabular}{cccc}
\hline
\hline
Observational interval & Year & $\tau_{\rm X-ray}$ & $\tau_{opt}$ \\
MJD & & d & d \\
\hline
$50373 - 50378$ & 1996 & $0.73\pm0.20$& $2.00\pm0.02$\\
$51336 - 51350$ & 1999 & $1.15\pm0.15$& $2.40\pm0.02$\\
$51627 - 51643$ & 2000 & $1.42\pm0.10$& $2.45\pm0.02$\\
$51681 - 51691$ & 2000 & $1.40\pm0.10$& $2.30\pm0.02$\\
\hline
\end{tabular}
\end{center}
\end{table*}

\clearpage

\clearpage

\clearpage

\clearpage

\clearpage

\clearpage

\clearpage

\clearpage

\end{document}